\documentclass[12pt]{iopart}

\usepackage[dvipsnames]{xcolor}
\usepackage{amssymb}
\usepackage[caption=false]{subfig}
\usepackage{graphicx}
\usepackage{hyperref}       
\usepackage{iopams}
\expandafter\let\csname equation*\endcsname\relax
\expandafter\let\csname endequation*\endcsname\relax
\usepackage{physics}
\usepackage{tikz}
\usepackage{cleveref}
\usepackage{url}            
\usepackage[switch]{lineno}
\usepackage[normalem]{ulem}

\colorlet{green}{OliveGreen}

\newcommand\boldmin{\mathbf{min}}
\newcommand\boldmax{\mathbf{max}}

\newcommand{\CC}{C\nolinebreak\hspace{-.05em}\raisebox{.4ex}{\tiny\bf +}\nolinebreak\hspace{-.10em}\raisebox{.4ex}{\tiny\bf +}}
\crefname{equation}{}{}
\crefname{figure}{}{}
\crefname{enumi}{}{}
\creflabelformat{equation}{#2#1#3}

\hypersetup{
pdftitle={Larmor Power Limit for Cyclotron Radiation of Relativistic Particles in a Waveguide.},
pdfsubject={physics.comp-ph, physics.ins-det, physics.class-ph},
pdfauthor={R.J.~Taylor, N.~Buzinsky},
pdfkeywords={CRES, cyclotron radiation, waveguide},
}

\begin{document}

\title{Larmor Power Limit for Cyclotron Radiation of Relativistic Particles in a Waveguide }
\author{N.~Buzinsky$^{1,2}$, R.\,J.~Taylor$^{3,4}$, W.~Byron$^{1,2}$, W.~DeGraw$^{1,2}$, B.~Dodson$^{1,2}$, M.~Fertl$^5$, A.~Garc\'{\i}a$^{1,2}$, A.\,P.~Goodson$^{2,6}$, B.~Graner$^{1,2}$, H.~Harrington$^{1,2}$, L.~Hayen$^{3,4}$, L.~Malavasi$^7$, D.~McClain$^{8,9}$, D.~Melconian$^{8,9}$, P.~M\"uller$^{10}$, E.~Novitski$^{1,2}$, N.\,S.~Oblath$^{11}$, R.\,G.\,H.~Robertson$^{1,2}$, G.~Rybka$^{1,2}$, G.~Savard$^{10}$, E.~Smith$^{1,2}$, D.\,D.~Stancil$^{12}$, D.\,W.~Storm$^{1,2}$, H.\,E.~Swanson$^{1,2}$, J.\,R.~Tedeschi$^{11}$, B.\,A.~VanDevender$^{1,11}$, F.\,E.~Wietfeldt$^{13}$, A.\,R.~Young$^{3,4}$}


\address{$^1$Department of Physics, University of Washington, Seattle, WA 98195}
\address{$^2$Center for Nuclear Physics and Astrophysics, University of Washington, Seattle, WA 98195}
\address{$^3$Physics Department, North Carolina State University, Raleigh, NC 27695} 
\address{$^4$The Triangle Universities Nuclear Laboratory, Durham, NC 27708}
\address{$^5$Institute for Physics, Johannes Gutenberg University Mainz, 55128 Mainz, Germany}
\address{$^6$Department of Electrical and Computer Engineering, University of Washington, Seattle, WA 98195}
\address{$^7$Department of Physics, Worcester Polytechnic Institute, Worcester, MA 01609}
\address{$^8$Department of Physics \& Astronomy, Texas A\&M University, College Station, TX 77843}
\address{$^9$Cyclotron Institute, Texas A\&M University, College Station, TX 77843}
\address{$^{10}$Physics Division, Argonne National Laboratory, 9700 S. Cass Ave., Argonne, IL 60439}
\address{$^{11}$Pacific Northwest National Laboratory, Richland, WA 99352}
\address{$^{12}$Department of Electrical and Computer Engineering, North Carolina State University, Raleigh, NC 27695}
\address{$^{13}$Department of Physics and Engineering Physics, Tulane University, New Orleans, LA 70118}

\eads{\mailto{nbuzin@uw.edu}, \mailto{rjtaylo2@ncsu.edu}}



\date{\today}
\begin{abstract}
Cyclotron radiation emission spectroscopy (CRES) is a modern technique for high-precision energy spectroscopy, in which the energy of a charged particle in a magnetic field is measured via the frequency of the emitted cyclotron radiation.
The He6-CRES collaboration aims to use CRES to probe beyond the standard model physics at the TeV scale by performing high-resolution and low-background beta-decay spectroscopy of ${}^6\textrm{He}$ and ${}^{19}\textrm{Ne}$. Having demonstrated the first observation of individual, high-energy (0.1 -- 2.5 MeV) positrons and electrons via their cyclotron radiation, the experiment provides a novel window into the radiation of relativistic charged particles in a waveguide via the time-derivative (slope) of the cyclotron radiation frequency, $\dd{f}_\textrm{c}/\dd{t}$. 
We show that analytic predictions for the total cyclotron radiation power emitted by a charged particle in circular and rectangular waveguides are approximately consistent with the Larmor formula, each scaling with the Lorentz factor of the underlying $e^\pm$ as $\gamma^4$.
This hypothesis is corroborated with experimental CRES slope data.
\end{abstract}



\section{Introduction}

Cyclotron Radiation Emission Spectroscopy (CRES)~\cite{artMonrealOriginal}, originally developed by the Project 8 collaboration for measuring the absolute mass scale of the neutrino from beta-decay endpoint spectroscopy of tritium~\cite{PhysRevLett.131.102502,artDeterminingNuMassP8}, uses cyclotron radiation as a high-precision probe of the energy of a radiating charged particle via the cyclotron frequency:

\begin{equation}
    \omega_\textrm{c} = \frac{q B}{m \gamma}= \frac{q B c^2}{E},
    \label{eqn:cyc_freq}
\end{equation}
for a particle with charge magnitude $q$ and mass $m$ in a uniform magnetic field with magnitude $B$. 
A measurement of the frequency of the emitted cyclotron radiation within a known magnetic field therefore determines the Lorentz factor $\gamma$, and as a result, the total energy $E = \gamma m c^2$ of the radiating particle. 

\begin{figure}[t]
    \centering
    \includegraphics[width=\textwidth]{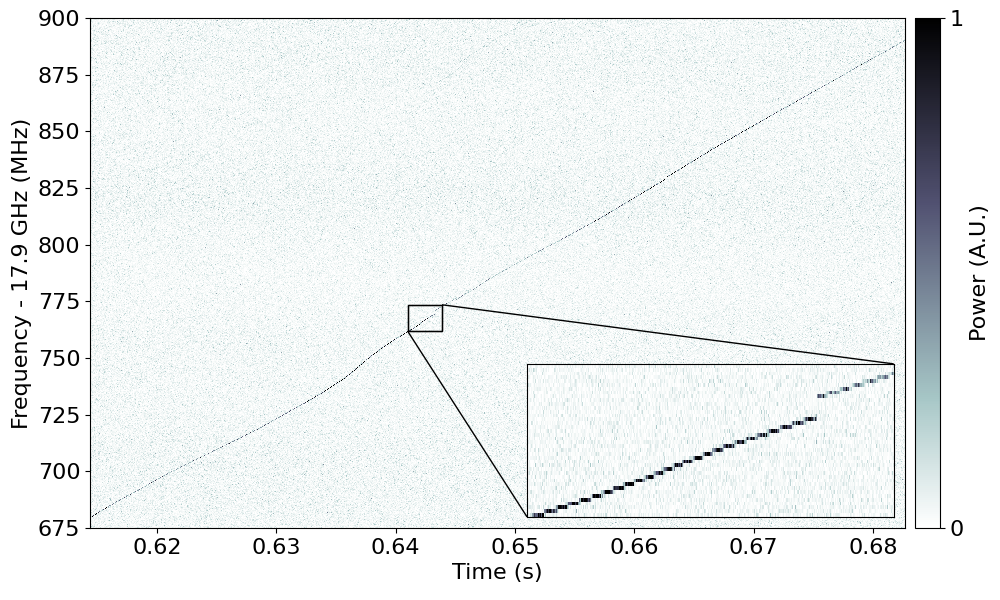}
    \caption{Time-frequency representation of a CRES event computed with the short-time Fourier transform (time bin width $\approx 6.83 \,\mu$s and frequency bin width $\approx 0.3$ MHz) for a beta decay of ${}^{19}\mathrm{Ne}$. Color corresponds to Fourier bin power (arbitrary units). For CRES, unbroken and approximately colinear sequences of high-power Fourier bins (tracks) are separated by frequency jumps corresponding to scatters of the emitting $e^\pm$ on residual gas. Track slopes are proportional to the total radiated power emitted by the $e^\pm$.}
    \label{fig:sample_event}
\end{figure}

Figure \ref{fig:sample_event} shows the time-frequency representation of a CRES \textit{event} resulting from the radiation signal from a single positron ($e^+$). Radio-frequency (RF) signals are amplified, digitized, and then converted to time-frequency space via the short-time Fourier transform, in which the Fourier transform is sequentially applied to 3.41 $\mu$s-long data segments, resulting in 0.3 MHz-wide frequency bins.
Due to a network bottleneck limiting our data transfer rate, we record every other frequency spectra, which are displayed as 6.83 $\mu$s long for visualization.
An event can be composed of any number of \textit{tracks}, which are consecutive and approximately colinear sequences of high-power Fourier bins.
Observed jump discontinuities in the instantaneous cyclotron frequency result from scattering between the $e^\pm$ and residual gas.
$e^\pm$s are confined axially within the experimental apparatus via a magnetic bottle trap and exit the trapping volume either by scattering or by turning off the magnetic trap, ending the corresponding CRES event.

Tracks are positively-sloped in time-frequency space because the cyclotron radiation decreases the energy of the emitting charged particle.
The time-derivative of \ref{eqn:cyc_freq} implies

\begin{equation}
    \frac{1}{\omega_\textrm{c}}\dv{\omega_\textrm{c}}{t} = \frac{1}{f_\textrm{c}}\dv{f_\textrm{c}}{t} =  - \frac{1}{E} \dv{E}{t}. 
    \label{eqn:slope_vs_power}
\end{equation}

Given negligible electric fields, CRES slopes are dominated by energy losses from cyclotron radiation. 
While CRES was originally conceived as a technique for high-precision energy spectroscopy via the measurement of the event start frequency, CRES also provides an additional experimental observable, namely the track slope, for sensitive studies of the energy-loss processes of charged particles. 
Slope data are robust to the total power radiated by the $e^\pm$, given \ref{eqn:slope_vs_power}.
By contrast, direct observations of Fourier bin powers provide an underestimate of the total radiated power since CRES signals can be distributed among multiple different frequency components (sidebands), some of which may evade detection due to limited detector bandwidth~\cite{thesisFurse,artPhenoPaper}.

Here we consider the application of CRES to the measurement of the total cyclotron radiation rate via the track slopes from high-energy electrons and positrons (0.1 -- 2.5 MeV) using the He6-CRES experimental apparatus~\cite{artPaper01,thesisByron}.
He6-CRES applies CRES to search for deviations of the ${}^{6} \textrm{He}$ and ${}^{19} \textrm{Ne}$ beta spectra from the predictions of the standard model of particle physics.
The deviations that would be expected from beyond the standard model (BSM) physics at energies above the TeV scale could show up in beta decay effectively as chirality-flipping tensor and scalar currents~\cite{cirigliano:2013}.

The apparatus consists of a 1.156-cm-diameter circular copper waveguide inside a superconducting solenoid with magnetic field uniformity on the order of 10 parts per million per cm.
Radioactive isotopes produced by the FN-tandem accelerator at the University of Washington~\cite{artCENPAVdG,KNECHT201143} are transported approximately 10 m, cleaned of contaminant gases by cryogenic and getter pumps, and compressed by a turbo pump into the waveguide interior.
A pair of RF-transparent Kapton windows confine the radioactive isotopes to a 14.6 cm-long region within the circular waveguide.
Cyclotron radiation originating from this volume is read out in one direction with a low-noise amplifier and absorbed in the other direction by a graphite-epoxy conical termination.
The energy spectra of ${}^{6}\textrm{He}$ and ${}^{19}\textrm{Ne}$ were observed for kinetic energies between $0.1-2.2$~MeV by varying the magnetic field magnitude in 0.25-T intervals between 0.75 -- 3.25 T in accordance with \ref{eqn:cyc_freq}, given the RF bandwidth of 17.9 -- 19.1 GHz.
Measurements of these spectra by He6-CRES currently set the high-energy frontier for the technique~\cite{artPaper01}, surpassing the 32.1 keV N-line conversion electrons from ${}^{83\textrm{m}}\textrm{Kr}$ observed by Project 8~\cite{artP82015PRL}.

Measurements of track slopes (\ref{eqn:slope_vs_power}) can be used to assess potential systematic uncertainties in the search for BSM physics with He6-CRES.
The signal-to-noise ratio (SNR), for example, depends directly on the track slope for a given Fourier window size, which has consequences for the resulting detection efficiency.
Track slopes can also give indirect information regarding the underlying kinematic parameters of the radiating $e^\pm$ (e.g. energy, axial velocity, and guiding center position) which can be used to improve energy resolution~\cite{thesisAliAshtari}.
Characterizing the experimental detection efficiency and energy resolution depends on accurate simulations of CRES signals, necessitating precise predictions for the total radiated power.

In section \ref{sec:math_review}, we begin by summarizing the power radiated by an $e^\pm$ in a waveguide, which we analytically evaluate for the circular waveguide in section \ref{sec:circular}.
Additional corrections to the calculation in section \ref{sec:circular} for realistic CRES detectors are evaluated in section \ref{sec:corrections}.
Section \ref{sec:software} then outlines details regarding the software implementation of the numerical cyclotron power calculations.
In section \ref{sec:results}, we then compare the output of these numerical cyclotron radiation power calculations with experimental CRES slope data, discussing the observed phenomena.

\section{Radiated Power from a Charged Particle in a Waveguide in the Presence of a Magnetic Field}\label{sec:powerrad}

\subsection{Mathematical Review} \label{sec:math_review}

A broad literature exists on the power emitted by a charged particle in a waveguide~\cite{thesisFurse,artPhenoPaper,artCollin,artBarkusov1987,artDokuchaev2001,10.1063/1.1703815,artKotanjyan,bookButs,PhysRevLett.47.233,PhysRevLett.97.030801,artHannekePRA,PhysRevA.83.052122}. 
The high-energy asymptotic behavior of the total radiated power from a relativistic charged particle undergoing cyclotron radiation in a waveguide has not been addressed in the literature, to our knowledge.
It is typical to invoke a simplifying symmetry, for instance, by considering the radiation emitted by a charged particle with its guiding center position along the central axis of a circular waveguide~\cite{artDokuchaev2001}.
In contrast to previous calculations~\cite{thesisFurse,artPhenoPaper,artCollin,artBarkusov1987,artDokuchaev2001,10.1063/1.1703815,bookButs,artHannekePRA,PhysRevA.83.052122}, we restrict our discussion to particles in circular motion, eliminating complications associated with axial motion (e.g. Doppler shifting) that are independent from the overall kinematic power scaling.
This simplification is well-suited to the main limitation of CRES: a detection efficiency that decreases rapidly as the axial velocity exceeds $\approx 1\%$ of the total velocity. 

Extending the notation from~\cite{artPhenoPaper,bookJackson,bookPozar},
the total field radiated by a charged particle in a longitudinally-uniform and lossless waveguide in the $\pm \vu{z}$ directions is a weighted sum over the complete propagating orthogonal basis of the waveguide field solutions:

\begin{align}
    \vb{E}^\pm(\vb{r},\omega) = \sum_\lambda  A^\pm_\lambda(\omega) \vb{E}^\pm_\lambda(\vb{r}, \omega) \label{eqn:E_total} \\
    \vb{H}^\pm(\vb{r},\omega) = \sum_\lambda  A^\pm_\lambda(\omega) \vb{H}^\pm_\lambda(\vb{r}, \omega). \label{eqn:H_total}
\end{align}
Here $\lambda$ is a combined index including eigenvalue indices $(n,m)$, mode basis $\chi$, and transverse electric/magnetic (TE/TM) boundary conditions, required to fully describe all orthogonal solutions. 
We define an orthogonal basis for the electromagnetic fields such that
\begin{equation}
    \int_\mathcal{A} \vb{E}^\pm_\lambda(\vb{r},\omega) \cdot \vb{E}^{\pm *}_{\lambda'}(\vb{r},\omega) \dd{A} =  0, \quad \textrm{for} \, \lambda \neq \lambda'.   \label{eqn:E_norm}
\end{equation}

Equation \ref{eqn:E_norm} therefore defines the transmitted power ($\mathcal{P}_\lambda$) over the  vector basis:
\begin{equation}
       \int_\mathcal{A}   \left[ \vb{E}^\pm_\lambda(\vb{r},\omega) \times \vb{H}^{\pm *}_{\lambda'}(\vb{r},\omega) \right] \cdot (\pm \vu{z}) \dd{A} = \mathcal{P}_\lambda \delta_{\lambda,\lambda'}, \label{eqn:unnorm}
\end{equation}
where $\delta_{\lambda,\lambda'}$ is the Kronecker delta symbol.

The total power emitted in a waveguide is therefore found by computing the Poynting flux using \cref{eqn:E_total,eqn:H_total} over the waveguide cross-section $\mathcal{A}$: 

\begin{equation}
    P^\pm(\omega) = \int_\mathcal{A} \left[ \vb{E}^\pm(\vb{r},\omega) \times \vb{H}^{\pm *}(\vb{r},\omega) \right] \cdot (\pm \vu{z}) \dd{A} \\ = \sum_{\lambda} \mathcal{P}_\lambda \abs{A_{\lambda}^\pm (\omega)}^2 \label{eqn:P_norm2}
\end{equation}
where by the Lorentz reciprocity theorem

\begin{equation}
A_{\lambda}^\pm (\omega) = -\frac{1}{2 \mathcal{P}_\lambda} \int_\mathcal{V} \vb{J}(\vb{r}, \omega) \cdot \vb{E}_{\lambda}^\mp (\vb{r},\omega) \dd{V}, \label{eqn:intA_EJ}
\end{equation}
and where $\vb{J}(\vb{r}, \omega)$ is the Fourier transform of the $e^\pm$ current density: 

\begin{equation}
\vb{J}(\vb{r}, \omega) = \lim_{T \to \infty} \frac{1}{T} \int_{0}^T \vb{J}(\vb{r},t) e^{-\mathrm{i}\omega t} \dd{t}. \label{eqn:J_FFT}
\end{equation}

The field expansions \cref{eqn:E_total,eqn:H_total} imply that we can multiply $\vb{E}^\pm_\lambda(\vb{r},\omega), \vb{H}^\pm_\lambda(\vb{r},\omega)$, and $1/A^\pm_\lambda(\omega)$ by an arbitrary factor for each mode without affecting the total fields $\vb{E}^\pm(\vb{r},\omega)$ and $\vb{H}^\pm(\vb{r},\omega)$.
Equations \cref{eqn:unnorm,eqn:P_norm2,eqn:intA_EJ} show that the total radiated power calculation is also invariant under this mode-dependent transformation via the normalization constant $\mathcal{P}_\lambda$.
For mathematical convenience, we use this mode-wise degree of freedom to opt for unitless electric fields in intermediary calculations over more standard (and cumbersome) unit conventions.

For a point particle with charge $q$ with position $\vb{r}_0(t)$ and velocity $\vb{v}(t)$, the current density is given by $\vb{J}(\vb{r},t) = q \vb{v}(t) \delta^3(\vb{r} - \vb{r}_0(t))$.
Substituting \ref{eqn:J_FFT} into \ref{eqn:intA_EJ} and evaluating the volume integral after swapping the order of integration implies

\begin{equation}
    A_{\lambda}^\pm (\omega) =  \lim_{T \to \infty} \frac{-1}{2 \mathcal{P}_\lambda T} \int_{0}^T q \vb{v}(t) \cdot \vb{E}_{\lambda}^\mp (\vb{r}_0(t),\omega)  e^{-\mathrm{i} \omega t} \dd{t} .
    \label{eqn:intA0_inf_lim}
\end{equation}

Furthermore, for a charged particle with zero axial velocity and with cyclotron period $T_\textrm{c} = 2 \pi / \omega_\textrm{c}$, the periodic integrand $\vb{v}(t) \cdot \vb{E}(t)$ of \ref{eqn:intA0_inf_lim}  has non-zero Fourier components only at integer multiples (\textit{harmonics}) of the cyclotron frequency.
In this case, \ref{eqn:intA0_inf_lim} reduces to

\begin{equation}
    A_{\lambda}^\pm (\omega) =  \frac{-1}{2 \mathcal{P}_\lambda T_c} \int_{0}^{T_c} q \vb{v}(t) \cdot \vb{E}_{\lambda}^\mp (\vb{r}_0(t),\omega)  e^{-\mathrm{i} \omega t} \dd{t} .
    \label{eqn:intA0}
\end{equation}

The total radiated power emitted by a charged particle in a waveguide is given by

\begin{equation}
    P_\textrm{tot}  = \sum_{\ell=1}^{\infty} \sum_{n,m}^{\infty}  P^\textrm{TE}_{n,m,\ell} + P^\textrm{TM}_{n,m,\ell} \label{eqn:final_sum0} \text{,}
\end{equation}
where
\begin{equation}
    P^\textrm{TE/TM}_{n,m,\ell} =  \sum_{\pm\vu{z}, \pm \ell, \chi}  \mathcal{P}_\lambda \abs{ A^\pm_{\lambda}(\ell \omega_c)}^2  \label{eqn:final_sum1}\text{,}
\end{equation}
in which we define $P^\textrm{TE/TM}_{n,m,\ell}$ to be the total power in the TE/TM $(n,m)$ mode summed over direction of propagation, mode basis $\chi$, and positive and negative harmonics $\omega = \pm \ell \omega_c$.
Variables $n,m,$ and $\ell$ are exclusively used to denote non-negative integers.
Zero-frequency $(\ell=0)$ fields do not propagate power. 

We note that \ref{eqn:final_sum1} is independent of the signs of $q$ and $\vb{v}(t)$, implying that the power radiated by a particle is identical to its antiparticle for all modes and harmonics, given identical guiding center positions.

Equations \cref{eqn:intA0,eqn:final_sum0,eqn:final_sum1} are relativistically valid, following from the Lorentz reciprocity theorem which itself is a consequence of Maxwell's equations~\cite{bookPozar}.
These results are expected to break down as the waveguide becomes effectively transparent in the limit of high photon energies due to finite values for the plasma frequency~\cite{bookAshcroftMermin} and the mass attenuation coefficient of the waveguide material~\cite{Gerward1982}. 
Quantum corrections to the radiation become significant when the median radiated photon has energy comparable to the particle rest mass~\cite{artTernov,bookSokolov}.
These effects occur well beyond the energy scales occurring in the He6-CRES experiment, so we do not consider them in the following derivation.

\subsection{Circular Waveguide} \label{sec:circular}
The specific case of cyclotron radiation in a circular waveguide is of particular experimental interest, given its simplicity.
Integration of \ref{eqn:intA0} for a radiating charged particle in a circular waveguide requires the evaluation of the product $\vb{v} \cdot \vb{E}$ over the particle trajectory. We consider a particle with fixed $z$ position, undergoing circular motion.
Using the notation illustrated in figure \ref{fig:geometry}, $\vb{v}$ and $\vb{E}$ are converted to a Cartesian coordinate basis as follows: 

\begin{figure}[t]
    \centering
    \includegraphics[width=0.75\columnwidth]{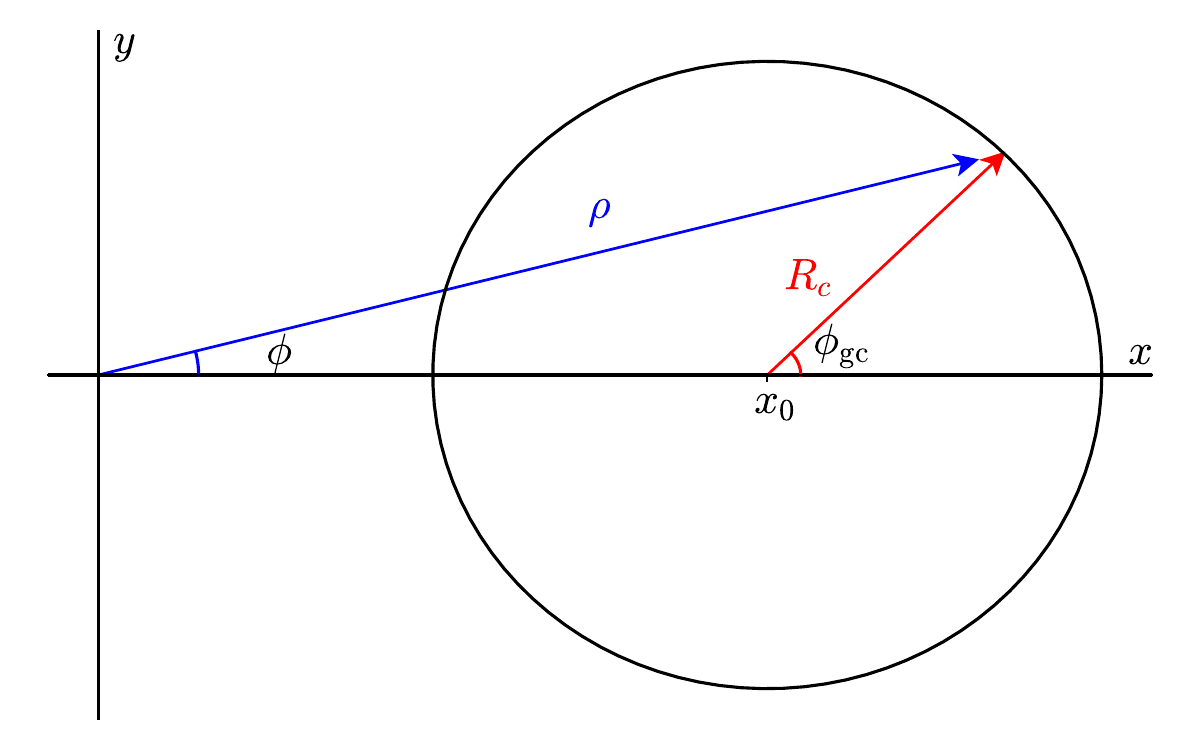}
    \caption{Notation used for cyclotron radiation power calculations of off-axis charged particles in a waveguide. While the mode fields $\vb{E}_\lambda$ are given in the coordinate system of the waveguide (circular: $(\rho,\phi)$, rectangular: $(x,y)$), it is convenient to integrate over $\phi_\textrm{gc}$, the angle of the $e^\pm$ with respect to its guiding center position.}
    \label{fig:geometry}
\end{figure}

\begin{equation}
 \vb{v}(\phi_\textrm{gc})  =  \begin{pmatrix}
v_x  \\
v_y 
\end{pmatrix}  = v \begin{pmatrix}
-\sin \phi_\textrm{gc} \\
+\cos \phi_\textrm{gc} 
\end{pmatrix}
\label{eqn:v_vec}
\end{equation}

\begin{equation}
\vb{E}  =  \begin{pmatrix}
E_x  \\
E_y 
\end{pmatrix}  =  \begin{pmatrix}
E_\rho \cos \phi - E_\phi \sin \phi \\
E_\rho \sin \phi + E_\phi \cos \phi 
\end{pmatrix} 
\end{equation}
where $\phi$, the azimuthal angle of the $e^\pm$ with respect to the central axis of the waveguide, is implicitly a function of $\phi_\textrm{gc}$, the azimuthal angle of the $e^\pm$ with respect to its guiding center position.

Changing variables from time to $\phi_\textrm{gc}$, \ref{eqn:intA0} then implies

\begin{equation}
    A^\pm_\lambda(\ell \omega_c) = - \frac{q v}{4 \pi \mathcal{P}_\lambda} \int_0^{2\pi} e^{-\mathrm{i}\ell\phi_\textrm{gc}}  \left[ -E^\mp_\rho \sin(\phi_\textrm{gc}-\phi) + E^\mp_\phi \cos(\phi_\textrm{gc}-\phi)  \right] \dd{\phi_\textrm{gc}}  .\label{eqn:A_dot_cyl}
\end{equation}

From~\cite{bookPozar}, the expressions for the transverse electric fields for TE and TM modes are:

\begin{equation}
     \vb{E}^\textrm{TE}_{\lambda}(\vb{r},\omega) = e^{-\mathrm{i}\beta_{n,m} z - \mathrm{i} n \chi \phi} \left[ - \frac{n}{k_{n,m} \rho} J_n(k_{n,m} \rho) \vu*{\rho} + \mathrm{i} \chi  J'_n(k_{n,m} \rho) \vu*{\phi} \right] \label{eqn:TE_circ}
\end{equation}
\begin{equation}
    \vb{E}^\textrm{TM}_{\lambda}(\vb{r},\omega) = e^{-\mathrm{i}\beta_{n,m} z - \mathrm{i} n \chi \phi} \left[ \mathrm{i} \chi  J'_n(k_{n,m} \rho)  \vu*{\rho} + \frac{ n}{k_{n,m} \rho} J_n(k_{n,m} \rho)  \vu*{\phi} \right] \label{eqn:TM_circ}
\end{equation}
where $J_{n}(x)$ is the $n$\textsuperscript{th}-order Bessel function of the first kind, $p^{(\prime)}_{n,m}$ is the m\textsuperscript{th} root of $J^{(\prime)}_{n}(x)$, and where $k_{n,m} = p^{(\prime)}_{n,m}/a$ is the TM (TE) cutoff wave number.
${\beta_{n,m}(\omega) = \sqrt{(\omega/c)^2 - k_{n,m}^2}}$ is the propagation constant characterizing the $z$-dependence of the mode field solutions.
$\chi = \pm 1$ corresponds to the $\phi$-basis of waveguide mode solutions within the circular waveguide. 

The analytic integral is made tractable via Graf's addition theorem~\cite{bookTISP}, represented in the coordinates of figure \ref{fig:geometry}: 

\begin{equation}
    e^{\pm \mathrm{i}n\phi} J_n(k\rho) = \sum_{u=-\infty}^{\infty} J_{n-u}(k x_0) J_u(k R_\textrm{c}) e^{ \pm \mathrm{i} u \phi_\textrm{gc}}. \label{eqn:graf}
\end{equation}

We recast \ref{eqn:A_dot_cyl} in terms of complex exponentials with Euler's formula and  \ref{eqn:TE_circ} in terms of $J_{n\pm 1}(k_{n,m} \rho)$ with the Bessel recurrence relations.
Applying Graf's addition theorem, the resulting integrals are of the form ${\int_0^{2\pi} e^{\mathrm{i} u \phi_\textrm{gc}} \dd{\phi_\textrm{gc}} = 2 \pi \delta_{u,0}}$, for arbitrary integer $u$. 
After summing over both $\phi$-polarizations, directions of propagation, and $\pm \ell \omega_c$,

\begin{equation}
P^\textrm{TE/TM}_{n,m,\ell} = \frac{\left( q v \right)^2}{2 \mathcal{P}_\lambda} \left[ J^2_{n+\ell}(k_{n,m} x_0) + J^2_{n-\ell}(k_{n,m} x_0) \right] 
     \begin{cases}
     J^{\prime 2}_\ell(k_{n,m} R_\textrm{c})  & \text{TE} \\
 \frac{\ell^2}{k_{n,m}^2 R_\textrm{c}^2}  J^{2}_\ell(k_{n,m} R_\textrm{c}) & \text{TM}
\end{cases} \label{eqn:circ_final}
\end{equation}
where

\begin{equation}
    \mathcal{P}_\lambda = \frac{\pi \epsilon_0}{2 k_{n,m}^2 } \begin{cases}  \frac{\beta_{n,m,\ell} c^2}{\abs{\ell} \omega_\textrm{c}}  \left( p_{n,m}^{\prime 2} - n^2 \right) J_{n}^ 2 (p^\prime_{n,m}) &\text{TE} \\
     \frac{\abs{\ell} \omega_\textrm{c}}{ \beta_{n,m,\ell}}  p_{n,m}^2 J_{n}^{\prime 2} (p_{n,m}) &\text{TM.}
    \end{cases} \label{eqn:circ_norm}
\end{equation}

Among the constants which set the scale of the radiated power, the propagation constant within the waveguide, $\beta_{n,m,\ell} \equiv \beta_{n,m}(\ell\omega_c)$, is particularly significant.
Since $\beta_{n,m,\ell}$ must be real for a given mode to propagate longitudinally along the waveguide, this constrains which modes and harmonics $(n,m,\ell)$ contribute to the overall power emitted by the particle. 

We claim that the cyclotron radiation power in a waveguide has a similar form regardless of the specific waveguide geometry. 
This is supported by \ref{app:rectangular} and \ref{app:general_proof}, which provide the corresponding derivations for the cyclotron radiation in a rectangular waveguide and in an arbitrary finite-area, longitudinally-uniform waveguide with a field basis derived from the Helmholtz equation.

\subsection{Corrections for Waveguide Resistivity and Termination Reflectivity} \label{sec:corrections}

\begin{figure}[t]
    \centering
    \includegraphics[width=\textwidth]{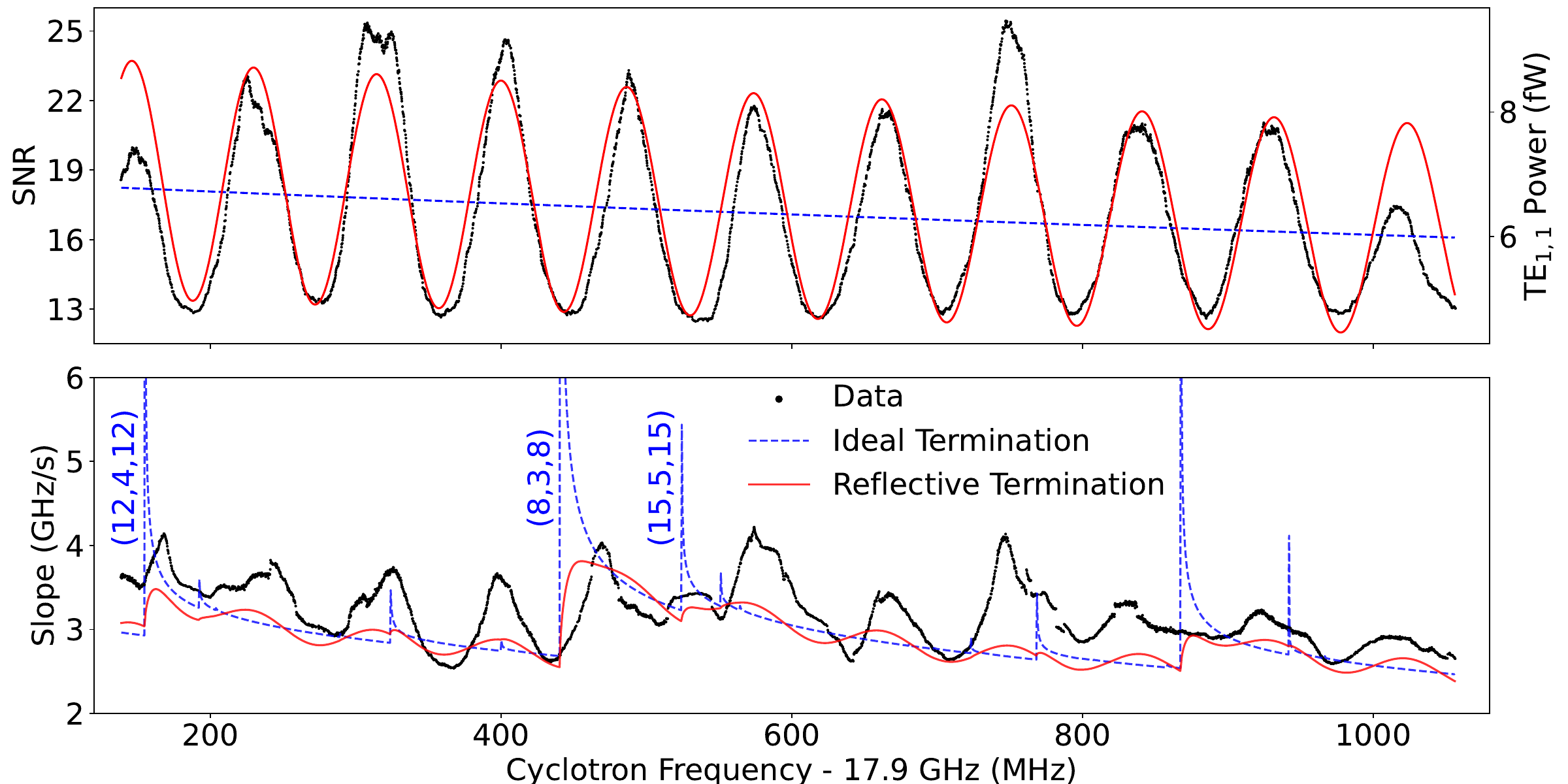}
    \caption{SNR (upper) and slope (lower) versus frequency for data (black) and simulations with (red) and without (dashed blue) accounting for termination reflections and finite waveguide conductivity. SNRs observed in data (upper, left axis) are compared to simulated $\mathrm{TE}_{1,1}$ powers (upper, right axis).
    SNR and slope data from the event from figure \ref{fig:sample_event}, observed at 1 T, are smoothed with a 70-Fourier bin-wide ($\approx$ 20 MHz) rectangular window, for clarity.
    The x-axis is given by the central frequency of the sliding rectangular window.
    In the ideal termination model, simulated slopes diverge at TE cutoff frequencies, several of which are labelled with their respective $n$, $m$, and $\ell$.}
    \label{fig:residuals}
\end{figure}

Real waveguide systems can have a variety of additional corrections, especially prominent at low energies~\cite{PhysRevLett.97.030801,PhysRevA.83.052122,project8collaboration2023cyclotron}, resulting in frequency-dependent modifications of the idealized model described in section \ref{sec:circular}.
Figure \ref{fig:residuals} shows how these discrepancies can significantly affect CRES event structure as a function of cyclotron frequency.
Data (black) shows the reconstructed high-power Fourier bins of the ${}^{19}\textrm{Ne}$ event illustrated in figure \ref{fig:sample_event}, observed at 1 T.
To reduce the effect of fluctuations, Fourier bins reconstructed as belonging to the event are assembled into a time-ordered list and smoothed with a 70-Fourier-bin-wide rectangular window ($\approx 20$ MHz). Data in figure \ref{fig:residuals} is plotted with respect to the frequency of the window center.
Slope data directly indicate the total power emitted by the radiating $e^\pm$, while SNR data indicate only the portion which is collected and amplified by the receiver chain.
Given the cutoff frequencies for the circular waveguide used for He6-CRES, $\mathrm{TE}_{1,1} \, (\ell=1)$ is the only non-evanescent (transmitting) mode within the detector bandwidth (17.9 -- 19.1 GHz) and is the sole contribution to the observed signal power.
The ideal termination model shows calculations corresponding to \cref{eqn:circ_final,eqn:circ_norm} for an on-axis $e^\pm$ in a perfectly terminated circular waveguide.
This minimal model (dashed blue) of the radiation evidently does not fully predict either SNR or slope data observed experimentally in He6-CRES.
The situation is remedied somewhat by extending this model to allow for reflections within the He6-CRES apparatus.
In particular, in figure \ref{fig:sample_event}, SNR variations can be observed with a period of approximately 90 MHz.
These variations are attributed to reflections off the He6-CRES termination, which did not perfectly absorb all incident radiation.
If the termination, at $z=D$ with respect to the $e^\pm$, has reflection coefficient $r$ with converse $t$ encompassing termination transmission and absorption such that energy conservation implies $r^2+t^2=1$, then the power radiated in the $\pm\hat{z}$ directions are no longer equal:

\begin{align}
  &P^{'+}_{n,m,\ell}  = \mathcal{P}_\lambda \sum_{\pm \ell, \chi} \abs{t A_{n,m}^+(\ell\omega_c)}^2  =  P_{n,m,\ell} \cdot \frac{t^2}{2} \label{eq:P_reflected_plus} \\
   &P^{'-}_{n,m,\ell}  = \mathcal{P}_\lambda \sum_{\pm \ell, \chi} \abs{A_{n,m}^-(\ell\omega_c) +  r e^{\mathrm{i}\Phi_{n,m,\ell}} A_{n,m}^+(\ell\omega_c) }^2 = P_{n,m,\ell} \cdot \left( \frac{1 + r^2}{2} +  r \cos\Phi_{n,m,\ell}\right)
\label{eq:P_reflected_minus}
\end{align}
where the mode-dependent round-trip phase distance is $\Phi_{n,m,\ell} = 2 \beta_{n,m,\ell} D $~\cite{artPhenoPaper}.
The total power is then
\begin{equation}
    P^{'}_{n,m,\ell} = P_{n,m,\ell} \cdot \left( 1 + r \cos \Phi_{n,m,\ell} \right).
    \label{eq:imperfectTerm}
\end{equation}

Observed SNRs depend on the $\mathrm{TE}_{1,1}\,(\ell=1)$ power radiated in the direction of the receiver chain (\ref{eq:P_reflected_minus}), while observed CRES slopes depend on the total power radiated in both directions (\ref{eq:imperfectTerm}).
In \cref{eq:P_reflected_plus,eq:P_reflected_minus,eq:imperfectTerm} we neglect the losses from an imperfectly conducting waveguide which would preferentially attenuate the reflected radiation over its longer path length.
Only the $\mathrm{TE}_{1,1}$ mode is assumed reflecting, given the observed SNR data, as a simplistic, lowest-order model for the He6-CRES apparatus. For the reflective termination model, the SNR data was fit to \ref{eq:imperfectTerm}, yielding $D=97.064(2)$ cm and $r=0.141(1)$.
The fit also provides an approximate calibration, mapping observed SNRs to radiated $\mathrm{TE}_{1,1}\, (\ell=1)$ powers.
The fit termination reflectivity is similar to frequency-dependent measurements taken at room temperature via vector network analyzer ($r\lesssim 0.14$) over the 17.9--19.1 GHz bandwidth.
However, the fit length is longer than the measured distance between the magnetic bottle trap center and the termination (90.4 cm).
A more complete model with all reflections and better fits to physical measurements is in active development.

Further consideration is needed to address the slope discontinuities that appear in the ideal termination model but not in data.
Slope discontinuities occur at frequencies corresponding to TE mode cutoffs, since the power emitted into a TE mode of a waveguide by a charged particle is inversely proportional to the propagation constant. 
If an $e^\pm$ has a cyclotron frequency satisfying $\omega_c = k_{n,m} c / \ell $ for integers $(n,m,\ell)$, then the power radiated into this mode will diverge.
Physically, this does not happen in the He6-CRES apparatus as the copper waveguide has non-zero resistivity, resulting in an imaginary attenuative term $\mathrm{i}\alpha_c$ for the propagation constant that removes this divergence. 
An approximation for the attenuation due to Ohmic losses in a circular waveguide can be derived for TE$_{n,m}$ and TM$_{n,m}$ modes for $\ell \omega_c> k_{n,m} c$ via the ratio of lost to incident power per unit length~\cite{bookBalanis}:

\begin{equation}
        \alpha_{c_\lambda} = \frac{R_s \ell \omega_c}{a\eta \beta_{n,m,\ell} c} \begin{cases}
            \left(\frac{k_{n,m} c}{\ell \omega_c} \right)^2 + \frac{n^2}{p_{n,m}^{\prime 2} - n^2} & \text{TE}\\
            1 & \text{TM}.
        \end{cases}\label{eqn:attenuation_balanis}
\end{equation}

Here, $\eta = \sqrt{\mu_0/\epsilon_0}$ is the impedance of free space and $R_s = \sqrt{\ell \omega_c\mu_0/2\sigma}$ is the surface resistance of the waveguide walls, where $\sigma$ is the electrical conductivity of the copper waveguide ($\approx 7.7 \times 10^9$ S/m at 35 K~\cite{10.1063/1.555614,bookEkin}).
This value for the conductivity of copper within the He6-CRES apparatus is validated by field dissipation predictions from~\cite{bookKnoepfel}, given experimental values of the waveguide wall thickness ($\approx 0.3$ mm) and the event end times, which persist for approximately 3.1 ms after turning off the magnetic bottle trap.

The addition of a physically motivated imaginary term to the propagation constant modifies the cutoff peaks of the model slopes in figure \ref{fig:residuals} to be more consistent with data.
Equation \ref{eqn:attenuation_balanis} is computed with respect to the infinite conductivity waveguide mode basis (\cref{eqn:TE_circ,eqn:TM_circ}). A finite conductivity waveguide modifies the boundary conditions and therefore the field basis $\vb{E}_\lambda, \vb{H}_\lambda$. 
More realistic models of finite conductivity waveguides near TE mode cutoffs are available~\cite{bookCollin}, though constructing the modified vector field basis puts this beyond the scope of the current work. These models are expected to result in smooth transitions between attenuated and propagating modes as a function of frequency, as observed in data (figure \ref{fig:residuals}, lower) but not simulations.

Including both $\mathrm{TE}_{1,1}$ termination reflections and resistive waveguide losses results in closer agreement between data and simulations (figure \ref{fig:residuals}).
The detailed frequency-dependent structure of CRES events is evidently significantly affected by the RF environment of the experimental apparatus.
A comprehensive characterization of the He6-CRES apparatus including other reflective surfaces, reflections of higher-order modes, and other physical deformations of the waveguides is currently incomplete, resulting in imperfect agreement between data and simulations.
Ongoing termination upgrades aim to minimize RF reflections within the He6-CRES apparatus, limiting both the SNR and slope variations that adversely affect the event reconstruction efficiency.

\subsection{Computational Costs and Software Implementation} \label{sec:software}
Computational cost is the most significant barrier to evaluating the total radiated power from a relativistic $e^\pm$ in a waveguide, given the large number of contributing modes.
It is convenient to form a lookup table for the Bessel roots and sum $\ell$ from $\lceil k_{n,m} c/ \omega_\textrm{c} \rceil$ to some maximum index $L$, where $f(x) = \lceil x \rceil$ is the ceiling function, guaranteeing that the mode propagates within the waveguide.

For a given harmonic, there are $\mathcal{O}(\ell^2)$ propagating modes in $(n,m)$-space~\cite{artNModes}, implying that \ref{eqn:final_sum0} has a total computational time cost of $\mathcal{O}(L^3)$ when evaluating up to a maximum harmonic $L$. Including many modes is critical for reducing truncation errors associated with the three-dimensional infinite sum, which results in underestimating the true value, given that all terms are positive. 

We note that when the charged particle guiding center position is along the central axis of the circular waveguide ($x_0 = 0$),  since $J_u(0) = \delta_{u,0}$ and $n$ and $\ell$ are both non-negative integers, $\abs{A_{n,m,\ell}}^2$ is only non-zero when $\ell=n$, reproducing the result in~\cite{artDokuchaev2001}. Notably, the computational cost for on-axis $e^\pm$s reduces to $\mathcal{O}(L^2)$ though these savings are largely inconsequential given that vanishingly few $e^\pm$s are produced on-axis due to the uniform source gas density within the decay cell of a CRES experiment.

Numerical evaluation of \ref{eqn:circ_final} is non-trivial, given the large number of propagating modes that contribute to the total radiated power emitted by an $e^\pm$ in a waveguide.
For these purposes, we developed \texttt{c-urchin} \footnote{\url{https://github.com/Helium6CRES/c-urchin}}, a lightweight implementation of the above calculations in \CC.
Time-costs for both circular and rectangular waveguides remain dominated by Bessel function evaluations. 
At $L = 100$, this corresponds to about $10^8$ Bessel function evaluations per $e^\pm$.
The analytical solutions have reduced the number of required Bessel function evaluations by approximately a factor of $10^3$ in comparison to direct numerical integration of \cref{eqn:intA0}, which otherwise provides a useful cross-check for the derived analytic expressions.
Caching and asymptotic approximations may be other potential avenues for reducing the computational resources needed per $e^\pm$.

Without analytic eigensolutions for the mode fields, such as for general waveguide geometries, or general charged particle trajectories arising from axial motion in a magnetic bottle trap, direct numerical integration of \cref{eqn:final_sum0,eqn:intA0} becomes prohibitively expensive. 

\section{Results and Discussion} \label{sec:results}
\subsection{Larmor Power Limit}

\begin{figure}[t]
    \centering
    \includegraphics[width=\textwidth]{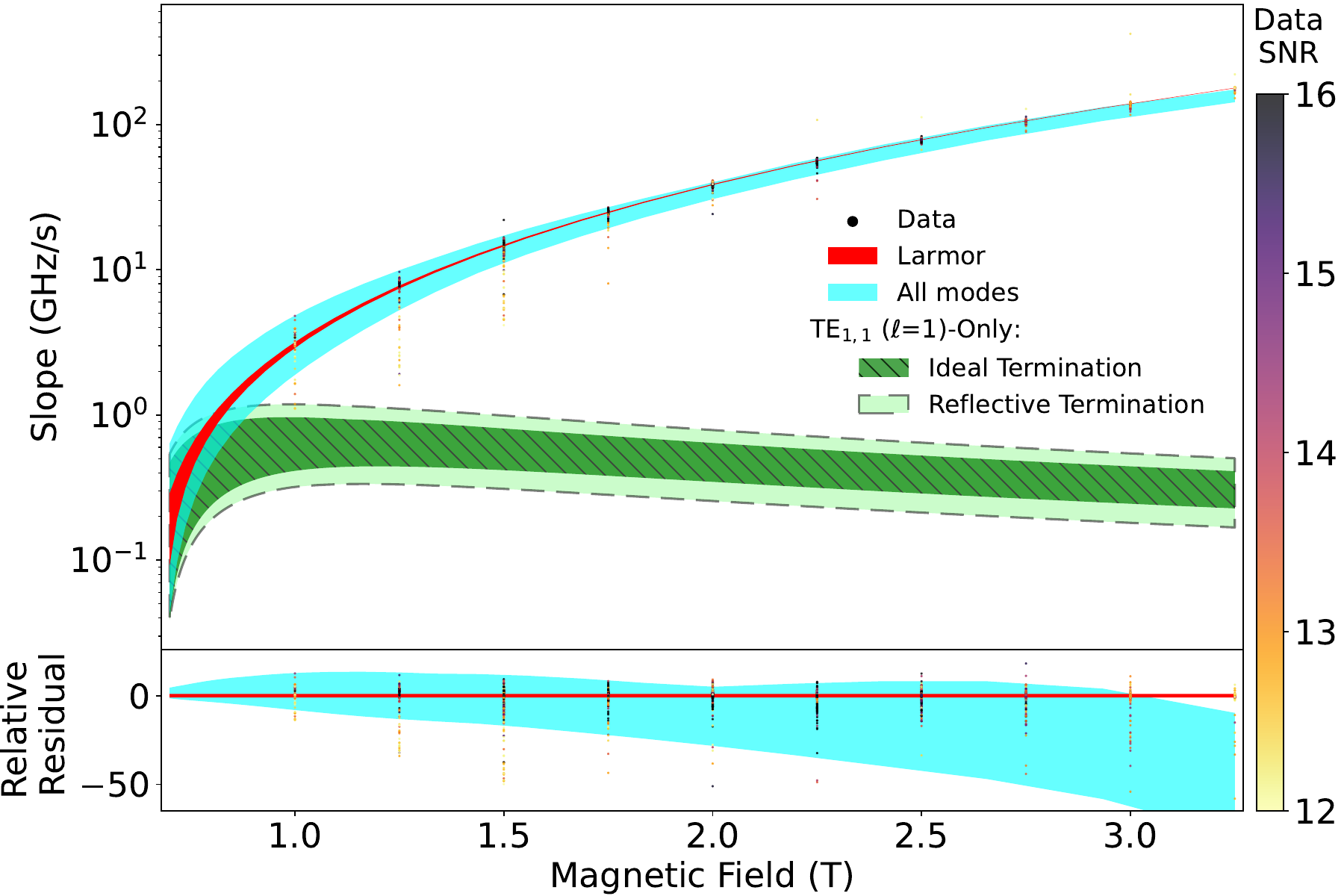}
    \caption{Observed track slopes, $\dd{f}_\textrm{c}/\dd{t}$, in comparison to the expectation from free-space (red) and from waveguide simulations in a circular waveguide (blue, green) as a function of experimental magnetic field. Data points show the average slopes of 2765 events from decays of ${}^{19}\textrm{Ne}$, colored by SNR (right). The colored bands correspond to the minimum/maximum slopes calculated over a grid of 12 guiding center positions and 23 cyclotron frequencies between 18 -- 19.1 GHz for a 35 K copper waveguide. Residuals are defined by a linear transformation (\ref{eqn:rel_residual}) such that the maximum and minimum Larmor slopes equal $\pm 1$, respectively.}
    \label{fig:slope_field}
\end{figure}

In figure \ref{fig:slope_field}, data points show the average (start-to-end) slopes of 2765 events observed from decays of ${}^{19}\textrm{Ne}$ as a function of magnetic field.
Data point color indicates the mean SNR of Fourier bins composing reconstructed events.
Slope and SNR variations are caused by differences in the underlying kinematic parameters (e.g. kinetic energy, guiding center position, axial velocity) of the radiating $e^\pm$.
Average event slopes are shown instead of more granular frequency-dependent data as they are less susceptible to priors in the event reconstruction algorithm~\cite{artPaper01,thesisChristine}, which have been tuned to initially search for events within a range of slopes, which includes the Larmor expectation.
The standard deviation of reconstructed event slopes are at least an order of magnitude smaller than the search window, indicating that the contribution of the prior to the reconstructed slope distribution is negligible.

Colored bands show the minimum/maximum slopes calculated over a coarse grid of $e^\pm$ energies and guiding center positions, found by mapping simulated power calculations (\cref{eqn:circ_final,eqn:circ_norm}) into equivalent track slopes via \ref{eqn:slope_vs_power} for either all transmitting modes (blue) or for just $\mathrm{TE}_{1,1}\, (\ell=1)$ (green).

The total radiated power within the circular waveguide is found to approach the Larmor power in the relativistic limit, given by~\cite{bookJackson}:

\begin{equation}
    P_\textrm{Larmor} = \frac{q^2 \omega_\textrm{c}^2 \gamma^4 v^2}{6 \pi \epsilon_0 c^3}.
    \label{eqn:larmorPowerBField}
\end{equation}

Relative residuals ($R$) are defined by the field-dependent slope transformation

\begin{equation}
R\left(\dv{f_c}{t} \right) = 2 \, \frac{\dv{f_c}{t} - \boldmin\left(\dv{f^\mathrm{Larmor}_c}{t} \right)} {\boldmax\left(\dv{f^\mathrm{Larmor}_c}{t}\right) - \boldmin\left( \dv{f^\mathrm{Larmor}_c}{t}\right)}- 1, \label{eqn:rel_residual}
\end{equation}
such that $\boldmax(), \boldmin()$ are evaluated at each field, resulting in residuals that equal $\pm 1$ for the maximum and minimum Larmor slopes, respectively (figure \ref{fig:slope_field}, lower).
Agreement between the Larmor power and CRES slope data is strongest for high-SNR events from $e^\pm$s with negligible axial velocities, as assumed in section \ref{sec:powerrad}.

The free-space correspondence is surprising given that the Liénard-Wiechert field solutions for a point particle in free-space are markedly different from the radiated field solutions in the waveguide, given the different electromagnetic boundary conditions between the two systems.
Qualitatively, free space corresponds to the cyclotron radiation propagating radially outwards (with infinite range), while in the waveguide, radiation propagates outwards axially.

CRES results are naturally contrasted to the crossed-dipole antenna, which consists of two perpendicular dipole antennas driven with signals 90$^\circ$ out of phase. 
The crossed-dipole antenna closely mimics the power, phase, and polarization distributions expected from CRES signals~\cite{AshtariEsfahani_2023} if the dipole arms are electrically small with respect to the wavelength of the emitted radiation.
In the limit of zero-length arms, the antenna becomes equivalent to the rotation of a single point-like electric dipole, and the radiation is emitted only at the fundamental frequency ($\ell=1$).
The strong parallels in the $\ell=1$ radiation for a CRES $e^\pm$ and a crossed-dipole antenna might lead one to incorrectly imagine that higher-order harmonics would likewise be similarly negligible.
However, as shown in figure \ref{fig:slope_field}, these higher-order harmonics dominate as the fractional contribution of $\mathrm{TE}_{1,1}$ ($\ell=1$) to the total radiated power approaches 0 for large $e^\pm$ energies. While the power radiated in $\mathrm{TE}_{1,1}$ ($\ell=1$) approaches a constant value for high $e^\pm$ energies, its contribution to the slope ($P^\textrm{TE}_{1,1,1} f_c / E$) approaches 0, resulting in the downward trend for the green band observed at high magnetic fields.
The downward trend in the blue band observed at high magnetic fields is attributed to computational limitations (truncation error, finite grid spacing), though the waveguide resistivity model could also be a contributing factor.

\begin{figure}[t]
    \centering
    \includegraphics[width=\textwidth]{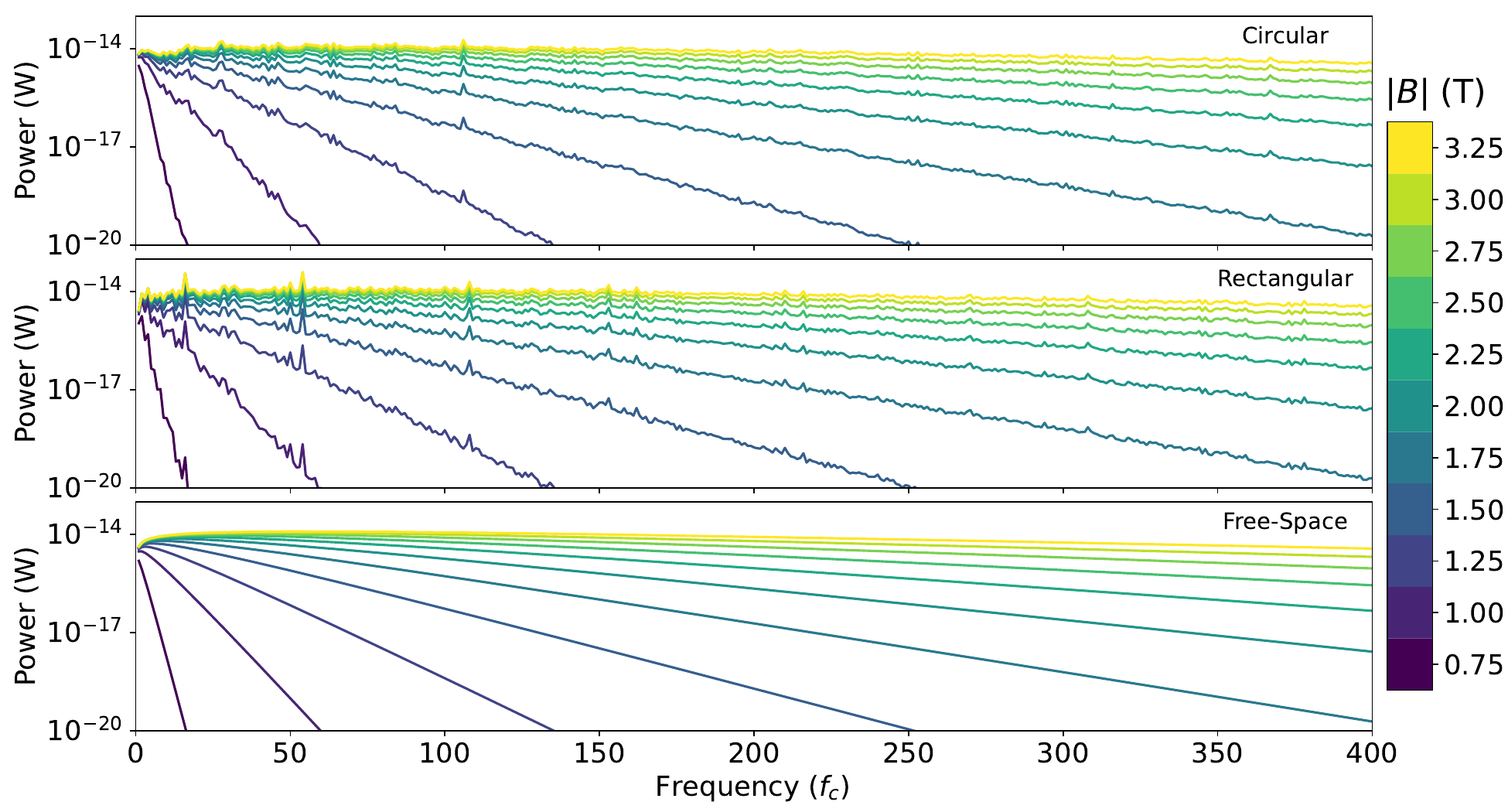}
    \caption{Power spectrum over the first 400 harmonic frequencies $\ell f_\textrm{c}$, as a function of magnetic field, for the circular waveguide (top), rectangular waveguide (center), and free-space (bottom), given $f_c$ = 18 GHz.}
    \label{fig:powerHarmonic}
\end{figure}

Figure \ref{fig:powerHarmonic} shows the predicted frequency distribution of emitted cyclotron radiation power from an $e^\pm$ with cyclotron frequency $f_c$ = 18 GHz, when in a circular waveguide (top), in a WR-42 rectangular waveguide (center), and in free-space~\cite{bookSchwinger} (bottom). 
$e^\pm$s are 2 mm off-axis in the waveguides, and the magnetic field is varied in 0.25 T steps from 0.75 -- 3.25 T, as a representation of typical He6-CRES events. 
 The resulting power spectra are approximately exponential at large harmonic numbers, with a decay constant dependent on the kinetic energy of the underlying charged particle. High energy $e^\pm$s require significantly more harmonics and modes in the numerical calculations for a given numerical error tolerance.
For a given kinetic energy, differences between the power spectra of $e^\pm$s are largely attributable to the specific TE resonances relatively close to the cyclotron frequency, which results in sharp peaks. 
Despite these cutoffs, which result in $\mathcal{O}(1)$ fractional deviations at particular harmonics from the free-space expectation, the numerical power spectra in the circular and rectangular waveguides are otherwise nearly identical to the free-space expectation.
Different guiding center positions or different waveguide sizes are found to change the location and magnitudes of these individual TE resonances, and not the asymptotic dependence on harmonic and $e^\pm$ energy.

\subsection{Discussion}

The analytic solution for the total radiated cyclotron power in a circular waveguide    (\cref{eqn:final_sum0,eqn:circ_final,eqn:circ_norm}) lacks the explicit $\gamma^4$-dependence present in the Larmor formula (\ref{eqn:larmorPowerBField}), despite the scaling being present in experimental CRES slope data and numerical sums of \cref{eqn:final_sum0,eqn:circ_final,eqn:circ_norm} (figure \ref{fig:slope_field}).
We note that in the relativistic limit ($v \to c$), the cyclotron radius ($R_c = v / \omega_c$)  approaches $c / \omega_c$. Despite minute differences in the $e^\pm$ trajectory, the total radiated power in a waveguide diverges in this limit, scaling with $\gamma^4$ as the summation of Bessel functions become approximately coherent~\cite{bookButs,bookSchott,bookWatson}.

We posit that the waveguide power calculations of \cref{eqn:circ_final} and \ref{app:rectangular}, summed over the eigensolutions in $(n,m)$-space, are approximately the Riemann sums for the corresponding Poynting vector integration in free space over $(\theta,\phi)$.

Specifically, the angular power distribution radiated at harmonic $\ell$ by a cyclotron particle in free-space is given by~\cite{bookSchwinger}

\begin{equation}
    \dv{P_\ell}{\Omega} = \frac{ \left(q \ell \omega_c v \right)^2}{8 \pi \epsilon_0 c^3} \left[ J^{\prime 2}_{\ell}\left( \frac{\ell v}{c} \sin \theta \right) + \left( \frac{J_\ell\left(\frac{\ell v}{c} \sin\theta \right)}{ \frac{v}{c} \tan\theta} \right)^2 \right]. \label{eqn:Ph_Schwinger_SA}
\end{equation}
If we consider the variable substitution

\begin{equation}
k' = \frac{\ell \omega_c}{c} \sin \theta,
\end{equation}
then \ref{eqn:Ph_Schwinger_SA} implies that in free space

\begin{equation}
    P_\ell = \int \dv{P_\ell}{\Omega} \dd{\Omega} = \frac{ q^2 v^2 \ell c }{4 \omega_c} \int_0^{\ell \omega_c / c} \left[ \mu_0 \omega_c^2 J^{\prime 2}_{\ell}(k' R_c) + \frac{\beta^{\prime 2}}{ \epsilon_0 \left(k' R_c \right)^2  } J^2_\ell(k'R_c) \right] \frac{k'}{\beta'} \dd{k'}, \label{eqn:k_int}
\end{equation}
where we used the identity $R_c = v / \omega_c$, and defined $\beta' \equiv \sqrt{(\ell \omega_c / c)^2 - k^{\prime 2}}$.
Equation \ref{eqn:k_int} evidently closely conforms to the general waveguide solutions from section \ref{app:general_proof}, which was evaluated explicitly for circular (\cref{eqn:circ_final,eqn:circ_norm}) and rectangular (\ref{app:rectangular}) geometries, barring the $\mathcal{O}(1)$ guiding-center position dependent factors.
Correspondence between the radiated cyclotron power emitted by an $e^\pm$ in a waveguide and in free space is therefore seen as resulting from the continuum limit of the sums over $k_{n,m}$-space. 
A similar relationship between  a square waveguide and free space has been shown for monopole antenna excitations~\cite{1330607}. This strongly suggests that this is a general relationship between radiation in a waveguide and free space.

\section{Conclusion}

Cyclotron radiation offers a uniquely sensitive probe for high-resolution and low-background spectroscopy of individual charged particles.
In addition to its core scientific motivation as a BSM physics discovery machine, He6-CRES naturally provides a novel experimental window into the radiation of relativistic charged particles in a waveguide via the track slope, $\dd{f}_\textrm{c}/\dd{t}$. We observe a close correspondence between the total radiated power emitted by individual $e^\pm$s in a circular waveguide and the free-space expectation given by the Larmor formula.
Simulations using the \texttt{c-urchin} software package, derived from first-principle analytic solutions for the power radiated by a relativistic particle in circular and rectangular waveguide geometries, provide further support to the claimed correspondence.
The analytic and computational advances described in this work greatly expands the ability of the He6-CRES collaboration to
better simulate and optimize future experimental upgrades.

\section*{Acknowledgments}

We thank Paul Kolbeck and Clint Wiseman for useful discussions.

This work is supported by the US Department of Energy (DOE), Office of Nuclear Physics, under Contracts No.~DE-AC02-06CH11357, DE-FG02-97ER41020, DE-FG02-ER41042, DE-AC05-76RL01830, DE-FG02-93ER40773, by the National Nuclear Security Administration under Award No.~DE-NA0003841, by the National Science Foundation, grants No. NSF-1914133 and PHY-2012395, by the Gordon and Betty Moore Foundation and by the Cluster of Excellence “Precision Physics, Fundamental Interactions, and Structure of Matter” (PRISMA+ EXC 2118/1) funded by the German Research Foundation (DFG) within the German Excellence Strategy (Project ID 39083149). The $^{83}{\rm Rb}-^{83}{\rm Kr}$ source used in this research was supplied by the DOE by the Isotope Program in the Office of Nuclear Physics.

\bibliographystyle{iopart-num}
\bibliography{references}  

\appendix
\section{Cyclotron Power Radiated by a Charged Particle in a Rectangular Waveguide}
\label{app:rectangular}

For a rectangular waveguide with dimensions $(a \times b)$ in the $x$ and $y$ directions, respectively, we evaluate \ref{eqn:intA0} directly in Cartesian coordinates. The loop integral is computed with respect to $\phi_\textrm{gc}$ as in figure \ref{fig:geometry}.
The position of the $e^\pm$ is notated:
\begin{align}
    \vb{r}_0(\phi_\textrm{gc}) =  \begin{pmatrix}
x_0 + R_\textrm{c} \cos \phi_\textrm{gc} \\
y_0 + R_\textrm{c} \sin \phi_\textrm{gc} 
\end{pmatrix}
\end{align}
where the most general guiding center position of the $e^\pm$ is $(x_0, y_0)$.

Utilizing the TE/TM electric fields:
\begin{equation}
     \vb{E}^\textrm{TE}_{n,m}(\vb{r},\omega) = \frac{\pi}{k_{n,m}}\left[  \frac{n}{b} \cos \frac{m \pi x}{a} \sin \frac{n \pi y}{b} \vu{x} -  \frac{m}{a} \sin \frac{m \pi x}{a} \cos \frac{n \pi y}{b} \vu{y} \right] \label{eqn:rect_E_TE}
\end{equation}
\begin{equation}
     \vb{E}^\textrm{TM}_{n,m}(\vb{r},\omega) = -\frac{\pi}{k_{n,m}}\left[  \frac{m}{a} \cos \frac{m \pi x}{a} \sin \frac{n \pi y}{b} \vu{x} + \frac{n}{b} \sin \frac{m \pi x}{a} \cos \frac{n \pi y}{b} \vu{y} \right] \label{eqn:rect_E_TM}
\end{equation}
we then define:
\begin{align}
\phi^\pm &= \phi^\pm_0 + \phi^\pm_{R_\textrm{c}} = \frac{m \pi x}{a} \pm \frac{n \pi y}{b} \\
\phi^\pm_0 &= \phi^x_0 \pm \phi^y_0  = \frac{m \pi x_0}{a} \pm \frac{n\pi y_0}{b} \\
\phi^\pm_{R_\textrm{c}} & = \frac{m \pi R_\textrm{c} }{a} \cos \phi_\textrm{gc} \pm \frac{n\pi R_\textrm{c}}{b} \sin \phi_\textrm{gc}.
\end{align}

The harmonic addition theorem then implies:
\begin{align}
& \phi^\pm_{R_\textrm{c}} =  k_{n,m} R_\textrm{c} \cos(\phi_\textrm{gc} \pm \varphi_{n,m}) \\
&\tan \varphi_{n,m}  = \frac{-n/b}{m/a}.
\end{align}

The trigonometric angle addition identities are then applied to separate out the constant $\phi^\pm_0$ terms. This results in 8 terms with integrals of the form:

\begin{align}
&\int_0^{2 \pi} e^{\mathrm{i} n u} \cos(z \cos(u + \varphi)) \dd{u} = \begin{cases}
   2 \pi \mathrm{i}^n e^{-\mathrm{i} n \varphi} J_n(z)  & n \text{ even} \\
  0 & n \text{ odd}
\end{cases} \label{eqn:bessel_identity_cos} \\
&\int_0^{2 \pi} e^{\mathrm{i} n u} \sin(z \cos(u + \varphi)) \dd{u} = \begin{cases}
  0  & n \text{ even} \\
2 \pi \mathrm{i}^{n-1} e^{-\mathrm{i} n \varphi} J_n(z) & n \text{ odd}
\end{cases} \label{eqn:bessel_identity_sin}
\end{align}
which are derived from the integral definition of the Bessel function $J_n(z) = \frac{1}{2\pi} \int_0^{2\pi} e^{\mathrm{i} n \theta} e^{-\mathrm{i} z \sin\theta} \dd{\theta}$.
Substituting in \cref{eqn:bessel_identity_cos,eqn:bessel_identity_sin} for the $\phi_\textrm{gc}$ integration and proceeding with the algebra yields:

\begin{align} 
P_{n,m,\ell}^\textrm{TE} &= \frac{q^2 v^2 }{4 \mathcal{P}_\lambda} J_\ell^{\prime 2}(k_{n,m} R_\textrm{c}) \left[ 1 + \cos 2\phi_0^y \cos 2 \ell\varphi_{n,m} + (-1)^\ell \cos2\phi_0^x \left( \cos 2\phi_0^y +  \cos 2 \ell\varphi_{n,m} \right)  \right]
 \label{eqn:rect_TE} \\
P_{n,m,\ell}^\textrm{TM} &= \frac{q^2 v^2}{ 4 \mathcal{P}_\lambda} \frac{\ell^2}{k^2_{n,m} R^2_\textrm{c} } J_\ell^{2}(k_{n,m} R_\textrm{c}) \left[ 1 - \cos 2\phi_0^y \cos 2 \ell\varphi_{n,m} + (-1)^\ell \cos2\phi_0^x \left( \cos 2\phi_0^y -  \cos 2 \ell\varphi_{n,m} \right)  \right] \label{eqn:rect_TM}
\end{align}

\begin{equation}
    \mathcal{P}_\lambda = \epsilon_0  \frac{a b}{4} \begin{cases} \frac{ c^2 \beta_{n,m,\ell}}{\abs{\ell} \omega_c}
    \left( 1 + \delta_{m,0} + \delta_{n,0} \right) &\text{TE} \\ 
    \frac{\abs{\ell} \omega_\textrm{c}}{ \beta_{n,m,\ell}} & \text{TM,} \end{cases} \label{eqn:rect_norm}
\end{equation}
where for TM both $n,m$ are positive integers, while for TE one of $n,m$ may equal 0. Disagreeing with~\cite{thesisFurse}, the guiding-center position dependence of \cref{eqn:rect_TE,eqn:rect_TM,eqn:rect_norm} is validated by direct numerical integration of \ref{eqn:intA0} for electric fields \cref{eqn:rect_E_TE,eqn:rect_E_TM}.

\section{Cyclotron Power Radiated by a Charged Particle in an Arbitrary Waveguide Geometry}
\label{app:general_proof}

For a longitudinally-uniform waveguide, separation of variables applied to the wave equation implies that the $H_z(x,y), E_z(x,y)$ fields satisfy the Helmholtz equation 

\begin{equation}
\pdv[2]{\psi_\lambda(x,y)}{x} + \pdv[2]{\psi_\lambda(x,y)}{y} + k_{\lambda}^2 \psi_\lambda(x,y) = 0,  \label{eqn:helmholtz}
\end{equation}

for TE and TM modes. Perfectly conductive waveguide walls result in Neumann (Dirichlet) boundary conditions.
Equation \ref{eqn:helmholtz} always has a countably infinite number of solutions $\psi_\lambda(x,y)$, which can be ordered by their (discrete) eigenvalue, $k_\lambda$~\cite{bookGilbarg}.
The transverse electric fields for mode $\lambda$ are derived from the solution to the Helmholtz equation:

\begin{align}
 &\vb{E}^\textrm{TE}_{\lambda}(\vb{r},\omega)=  \frac{1}{k_\lambda} \left[ 
  \pdv{\psi_\lambda(x,y)}{y} \vu{x}
  -\pdv{\psi_\lambda(x,y)}{x} \vu{y} \right] \label{eqn:TEvu_vec_deriv} \\
   &\vb{E}^\textrm{TM}_{\lambda}(\vb{r},\omega)=  \frac{1}{k_\lambda} \left[ 
  \pdv{\psi_\lambda(x,y)}{x} \vu{x}
  +\pdv{\psi_\lambda(x,y)}{y} \vu{y} \right].  \label{eqn:TMvu_vec_deriv}
\end{align}

Additionally, given that any function can be represented via its Fourier transform: 

\begin{equation}
    \psi_\lambda(x,y) = \int_{-\infty}^\infty \int_{-\infty}^\infty \widetilde{\psi}_\lambda(k_x,k_y) e^{\mathrm{i}k_x x + \mathrm{i}k_y y} \dd{k_x} \dd{k_y}, \label{eqn:helmholtz_Fourier_gen}
\end{equation}
general solutions to \ref{eqn:helmholtz} have non-zero Fourier components $\widetilde{\psi}_\lambda(k_x,k_y)$ only if $k_x^2 + k_y^2 = k_{\lambda}^2$, which is represented in polar coordinates of $k$-space as~\cite{bookMorseFeshbach}

\begin{equation}
    \psi_\lambda(x,y) = \int_0^{2\pi} \widetilde{\psi}_\lambda(\delta) e^{\mathrm{i}k_{\lambda} \left( x \cos\delta  + y \sin\delta  \right)} \dd{\delta}, \label{eqn:helmholtz_circ_k}
\end{equation}
where the waveguide is assumed to have a closed transverse cross-section.
Solutions to \ref{eqn:helmholtz} in arbitrarily-shaped waveguides are represented as a sum of plane waves, with wave number $k_\lambda$, over all directions. $\widetilde{\psi}_\lambda(\delta)$ encapsulates the phases and amplitudes of each plane wave contribution to the eigenmode $\lambda$.

The transverse fields \cref{eqn:TEvu_vec_deriv,eqn:TMvu_vec_deriv} are readily evaluated for the general solution \ref{eqn:helmholtz_circ_k} to the Helmholtz equation.
We rely on the smoothness of $\psi_\lambda(x,y)$, which is true for the Dirichlet boundary condition (TM) \cite{bookGilbarg}, in order to swap the orders of integration and differentiation.
The Neumann boundary condition (TE) requires smoothness of the boundary such that the normal derivative is defined almost everywhere.
Exotic waveguide geometries (e.g. Koch snowflake~\cite{Ziemann:2017nfp}), which are continuous, but nowhere-differentiable, are beyond the scope of this document as unphysical, with undefined TE modes.

Having expressions for the electric fields of the mode, the power radiated into mode $\lambda$ (\ref{eqn:final_sum1}) can be evaluated analogously to section \ref{sec:circular}, using  \cref{eqn:unnorm,eqn:intA0}.
Equation \ref{eqn:intA0} is a double integral over $\delta \times \phi_\textrm{gc}$-space, the latter of which can be integrated analytically by choosing the origin of our coordinate system to align with the guiding center position of the $e^\pm$, such that $x = R_c \cos\phi_\textrm{gc}$ and $y = R_c \sin\phi_\textrm{gc}$:

\begin{equation}
    P_{\lambda}(\ell \omega_c) = \frac{q^2 v^2}{2 \mathcal{P}_\lambda} \abs{\Psi_{\lambda,\ell}}^2  \begin{cases} \ell^2 J^{'2}_\ell(k_{\lambda}R_c)  & \text{TE} \\  \frac{\ell^2}{k^2_{\lambda} R_c^2} J_\ell^2(k_{\lambda} R_c) & \text{TM} \end{cases} \label{eqn:power_cases}
\end{equation}
where

\begin{equation}
      \Psi_{\lambda,\ell} = \int_0^{2\pi} \widetilde{\psi}_\lambda(\delta) e^{-\mathrm{i}\ell\delta} \dd{\delta},\label{eqn:gen_position_dependence}
\end{equation}
\begin{equation}
\mathcal{P}_\lambda(\ell \omega_c) = \epsilon_0 \int_\mathcal{A} \abs{\psi_\lambda(\vb{r})}^2 \dd{A} \cdot \begin{cases}
    \frac{c^2 \beta_\lambda(\ell\omega_c)}{\abs{\ell}\omega_c} & \text{TE}\\
     \frac{\abs{\ell}\omega_c}{\beta_\lambda(\ell\omega_c)} & \text{TM}.
\end{cases}\label{eqn:P_norm_general}
\end{equation}

If $\Psi_{\lambda,\ell}$ were completely arbitrary, the radiated power per mode $\lambda$ \ref{eqn:power_cases} would be likewise arbitrary.
In fact, since $\widetilde{\psi}_\lambda(\delta)$ represents the plane-wave expansion of the scalar solution to the Helmholtz equation in the waveguide, centered with respect to the guiding center position of the $e^\pm$, $\Psi_{\lambda,\ell}$ is independent of the kinematic parameters of the $e^\pm$ $(v,R_c)$. 
$\Psi_{\lambda,\ell}$ implicitly depends on the waveguide geometry, via the imposition of boundary conditions on $\widetilde{\psi}_\lambda(\delta)$, and the $e^\pm$ guiding center position, via the coordinate system. 
The power radiated into mode $\lambda$ by an $e^\pm$ in circular motion in circular (\cref{eqn:circ_final,eqn:circ_norm}) and rectangular (\ref{app:rectangular}) waveguide geometries
are consistent with the general expressions \cref{eqn:power_cases,eqn:gen_position_dependence,eqn:P_norm_general}, exhibiting the same factorization between geometric and kinematic terms.

\end{document}